\documentclass[5p,times]{elsarticle}

\usepackage{amsmath,amsfonts}
\usepackage{algorithmic}
\usepackage{algorithm}
\usepackage[caption=false,font=normalsize,labelfont=sf,textfont=sf]{subfig}
\usepackage{textcomp}
\usepackage{stfloats}
\usepackage{url}

\usepackage{graphicx}
\usepackage{comment}
\usepackage{color}
\usepackage{rotating} 
\usepackage{xcolor, colortbl}
\usepackage{lscape}
\usepackage{array, multirow, multicol}
\usepackage{amsmath,bm} 
\usepackage{url}
\usepackage{enumitem}
\newif\ifremark
\long\def\remark#1{
\ifremark%
        \begingroup%
        \dimen0=\columnwidth
        \advance\dimen0 by -1in%
        \setbox0=\hbox{\parbox[b]{\dimen0}{\protect\em #1}}
        \dimen1=\ht0\advance\dimen1 by 2pt%
        \dimen2=\dp0\advance\dimen2 by 2pt%
        \vskip 0.25pt%
        \hbox to \columnwidth{%
                \vrule height\dimen1 width 3pt depth\dimen2%
                \hss\copy0\hss%
                \vrule height\dimen1 width 3pt depth\dimen2%
        }%
        \endgroup%
\fi}

\remarktrue

\begin{document}
\begin{frontmatter}
\title{A New Family of Thread to Core Allocation Policies for an SMT ARM Processor}

\author[inst1]{Marta Navarro}

\affiliation[inst1]{organization={Universitat Politècnica de València},
            addressline={Camí de Vera s/n}, 
            city={València},
            postcode={46022}, 
            country={Spain}}

\author[inst1]{Josué Feliu}
\author[inst1]{Salvador Petit}
\author[inst1]{María E. Gómez}
\author[inst1]{Julio Sahuquillo}


\begin{abstract}
Modern high-performance servers commonly integrate Simultaneous Multithreading (SMT) processors, which efficiently boosts throughput over single-threaded cores. Optimizing performance in SMT processors faces challenges due to the inter-application interference within each SMT core. To mitigate the interference, thread-to-core (T2C) allocation policies play a pivotal role. State-of-the-art T2C policies work in two steps: i) building a per-application performance stack using performance counters and ii) building performance prediction models to identify the best pairs of applications to run on each core.

This paper explores distinct ways to build the performance stack in ARM processors and introduces the Instructions and Stalls Cycles (ISC) stack, a novel approach to overcome ARM PMU limitations. The ISC stacks are used as inputs for a performance prediction model to estimate the applications' performance considering the inter-application interference. The accuracy of the prediction model (second step) depends on the accuracy of the performance stack (first step); thus, the higher the accuracy of the performance stack, the higher the potential performance gains obtained by the T2C allocation policy. 

This paper presents SYNPA as a family of T2C allocation policies. Experimental results show that $SYNPA4$, the best-performing SYNPA variant, outperforms turnaround time by 38\% over Linux, which represents 3$\times$ the gains achieved by the state-of-the-art policies for ARM processors. Furthermore, the multiple discussions and refinements presented throughout this paper can be applied to other SMT processors from distinct vendors and are aimed at helping performance analysts build performance stacks for accurate performance estimates in real processors.

\end{abstract}


\end{frontmatter}

\section{Introduction}\label{sec:intro}

Modern high-performance servers implement simultaneous multithreading (SMT)~\cite{TUL95} cores as they increase the processor throughput with minimal area increase~\cite{ThunderX3}. SMT cores support the execution of multiple threads (or applications) that compete for the major core resources, including the issue logic, the execution units, and the (L1 instruction and L1 data) caches. In this way, the overall processor throughput is improved, but the inter-application interference rises, which translates into performance drops of individual applications over isolated execution. 
In other words, the performance of individual applications can be severely affected by inter-application interference. 
To mitigate the interference effects, the co-runner (i.e., the threads running in the same core) should be carefully chosen.

The previous rationale means that the performance of an SMT processor depends on the OS scheduler or, more precisely, the thread-to-core (T2C) allocation policy. Assuming a 2-thread SMT processor, this policy states the rules in which applications are paired and pinned to cores. The main aim of this policy is to reduce the intra-core interference among the selected pairs. In other words, to identify \emph{synergistic applications} as applications that exhibit complementary behaviors from a resource consumption perspective.

Designing T2C interference-aware allocation policies for a real processor needs to address important challenges. The first one is to determine if it is possible to characterize the \emph{synergy of applications} with the performance counters available in the target processor. The difficulty lies in that modern processors can capture hundreds of performance events that are gathered in a wide range of magnitudes (e.g., misses, stalls, cycles, etc.). The challenge comprises two major decisions: at which processor stage should performance be analyzed and which performance events should be considered.
Once the performance counters have been identified, an approach needs to be devised to characterize applications from a resource consumption perspective. 
In this paper, we use the \emph{Instructions and Stalls Cycles} (ISC) stack approach that helps identify which resources stalls come from.

In a previous work~\cite{ipdps-24}, we proposed SYNPA as an approach to a thread-to-core interference-aware policy for SMT processors. SYNPA uses the ISC stack based on three main categories for an ARM processor: \emph{backend} stalls, \emph{frontend} stalls, and \emph{dispatch} stalls.
The values of these categories are used to train a linear regression model to estimate the performance of an application depending on the co-runner. After evaluating the possible combinations, SYNPA selects and allocates the best pairs based on the estimated performance.
Because of the performance monitoring unit (PMU) constraints, SYNPA is built on top of an ISC stack that does not capture all the execution cycles, which introduces some measurement inaccuracies. 
Although the regression model trained with the experimental data hides these inaccuracies, it also leaves potential performance untapped.
More precisely, an ideal ISC stack should capture 100\% of the processor cycles; nevertheless, most PMUs existing on real processors do not since a wide set of performance events are not exclusive, but their counts can overlap each other.

In this work, we extend the original SYNPA approach in two main ways. First, we model the \emph{horizontal waste} (unused dispatch slots), which is particularly important for specific applications.
Second, due to the measured values do not account for the exact amount of processor cycles in a real processor, we propose and analyze different approaches to distribute the cycles and build a representative ISC stack. These improvements led to an extended version of SYNPA as a family of different versions using distinct stacks, where each member has three or four categories. Therefore, from now on, we will refer to SYNPA$3_{N}$ as the original SYNPA approach.

This paper discusses the authors' experience building the ISC stack from real performance counters, focusing on the alternatives devised to build the stack. In this regard, we dig deep into the PMU of ARM processors, discuss its limitations to build a stack that characterizes performance accurately and analyze different alternatives to overcome these limitations. 
Each alternative leads to a new T2C algorithm, obtaining a family of interference-aware schedulers.

This paper makes the following main contributions: 
\begin{itemize}
\item Introduces a new simple and effective approach, named the Instructions and Stalls Cycles (ISC) stack, to model the execution cycle distribution in SMT processors. 
\item A methodology is followed to build a family of stacks, called ISC$\textit{X}_{Y}$, considering different fields or categories and combining them to build the versions of SYNPA. 

\item The devised SMT scheduler, SYNPA, applies a performance prediction model using the ISC stack to estimate the applications to be executed in the same physical core.
\item Results show that when using SYNPA4, with only four categories, the SMT scheduler outperforms the Linux OS by 38\% in terms of turnaround time.

\end{itemize}

\section{Related work}

Some works on T2C allocation policies have focused on the use of heuristics to make scheduling decisions. 
In \cite{Snavely_symbiotic}, authors periodically run a subset of combinations to sample their performance, while in \cite{settle2004architectural}, offline profiling is used to improve processor throughput. However, neither of these two approaches is practical in a real scenario because they introduce non-negligible sampling and profiling overheads. Other approaches \cite{l1-aware,vega13,cal-marta} apply heuristics in real machines to guide the scheduling decisions with minimal overhead.

Other works closer to ours make scheduling decisions following a two-step approach. First, they develop a \emph{performance stack}, which puts on the same bar \emph{performance characteristics} of the running applications. Then, these stacks feed a performance predictor model to estimate the performance of pairs of applications if they were executed together on the same core. 

\textbf{Performance stacks.} Building a performance stack is a typical way to characterize the performance. Multiple ways
of constructing this stack are possible according to the used fields.
In ~\cite{Eyerman_CPIacc,Eyerman_SMTacc} seminal works were proposed from a theoretical perspective focusing on generic single-thread~\cite{Eyerman_CPIacc} and SMT~\cite{Eyerman_SMTacc} processors. 
Other works focus on real systems and deal with the constraints imposed by the PMUs of their target architectures. Yasin~\cite{top-down} builds a stack that characterizes performance in Intel processors at the issue stage, while Feliu et al.~\cite{symbiotic-josue} leverage the CPI accounting mechanisms of the IBM POWER8, which counts stalls at the commit stage, to compute the stacks that drive their symbiotic scheduler. 

Unlike previous work, in this paper, we devise the instruction and stall components (ISC) stack, a novel stack that characterizes distinct stall components at the dispatch stage available in the ARM processor, as well as speculatively executed instructions.
To the best of our knowledge, no previous work has focused on building performance stacks to guide scheduling decisions in ARM processors. In our work, we dig deep into the PMU of ARM processors and discuss its limitations. Multiple performance stacks are proposed in a refined way to improve the performance estimation model's accuracy, which will be discussed next.

\textbf{Performance prediction models.} 
Once applications have been characterized with a specific performance stack, a performance prediction model can be applied to estimate the performance that the distinct pairs of applications would have if they run together on the same core.
For instance, Moseley et al.~\cite{Moseley} employ linear modeling and recursive partitioning to estimate the speedup when executing two applications simultaneously. Radojkovic et al.~\cite{Radojkovic_2012} propose a statistical inference method to estimate the performance of the optimal task assignment. The use of performance models to guide scheduling is not limited to SMT processors. For example, In SMiTE~\cite{smite}, authors propose a regression model that estimates performance interference in the cache hierarchy.
Eyerman and Eeckhout~\cite{eyerman10} propose an analytical model that uses performance characterization stacks to predict the slowdown suffered by each application when co-scheduled with other applications. However, these performance stacks require custom hardware events and thus cannot be gathered in commercial processors. A similar approach but leveraging the CPI accounting mechanism of the IBM POWER8 was proposed by Feliu et al.~\cite{symbiotic-josue}. 

One of the main advantages of SYNPA~\cite{ipdps-24} compared to existing approaches is that thanks to using the ISC stacks proposed in this work, SYNPA uses fewer hardware events, which means it has significantly lower overhead than the models used in previous works without losing accuracy.

\section{The Challenge of Modeling the Performance} \label{sec:challenge}

Characterizing the behavior of the applications from a resource consumption perspective helps system hardware and software developers improve performance. For instance, it allows for identifying performance bottlenecks (e.g., insufficient memory bandwidth), which can later be relieved with modifications to the hardware (e.g., by adding an extra memory module) or software (e.g., improving the data organization). Regarding 
system software, it enables the implementation of intelligent scheduling algorithms that leverage characterizing the applications to develop performance prediction models that help improve the selection of co-running applications and their allocation to processor cores, minimizing the inter-application interference in the shared resources.

Nevertheless, characterizing the performance of applications for accurate performance estimates in a real processor is a complex task, among others, because the stack construction is limited by the processor's PMU, which defines the set of events that can be measured in the target architecture. Unfortunately, PMUs of commercial processors are far from ideal and introduce many constraints to obtain a precise performance characterization, as we explain below and demonstrate with experimental data.

\subsection{Problem Description: Ideal PMU}

The simplest way to expose the execution time of an application in a scalar processor is by presenting execution cycles as the sum of instructions and stalls (Equation \ref{eq:i+s}):

\begin{equation}
Cycles = Instructions + Stalls 
\label{eq:i+s}
\end{equation}

More precisely, instructions refer to the number of instructions that execute and make progress through a particular point in the pipeline; in other words, they represent useful work. On the contrary, stalls indicate cycles in which instructions are prevented from advancing in the pipeline due to data dependencies or structural hazards. Thus, stalls represent a lack of progress. There are countless sources of stalls in a processor. For example, a true data dependence stall appears when a given instruction depends on an older memory access instruction that has not written back its produced data yet; thus, the consuming instruction is kept waiting (stalled) until the required data becomes available. Similarly, a structural hazard is caused by the lack of availability of a particular resource (e.g., an ROB entry or an arithmetic operator) needed by a given instruction to make progress through the pipeline. Control hazards, such as branch mispredictions, can also be included in the stalls component. 

This simple example complicates in a superscalar processor. Let us assume a $n$-wide superscalar processor. Now, performance can be expressed by Equation \ref{eq:superscalar}:
\begin{equation}
Cycles = \frac{Instructions + Stalls}{n}  
\label{eq:superscalar}
\end{equation}

The terms of Equation \ref{eq:i+s} are now divided by $n$ since, in a given cycle, the sum of instructions and stalls is equal to $n$. For example, assuming that the stack is built at the issue stage in a 4-wide superscalar processor, if one instruction is issued, then the number of stalls would be 3. 

The complexity is further aggravated in an SMT processor, which is the focus of this paper. In an SMT processor, instructions from several applications can be executed in the same cycle, and the characterization needs to consider both instructions and stalls separately for each thread. The execution time of an application $ap\_{i}$, in a 2-way SMT 4-way superscalar processor running $ap\_{1}$ and $ap\_{2}$ can be expressed using Equation \ref{eq:smt-per-thread}:

\begin{equation}
Cycles_{ap\_{i}} = \frac{Instructions_{ap\_{i}}+ Stalls_{ap\_{i}}}{4} 
\label{eq:smt-per-thread}
\end{equation}

In this case, the simplest way to implement an accounting architecture for each thread is to consider all the cycles where that thread does not issue any instruction as stalls. In other words, we are modeling that the number of stalls rises in SMT mode due to co-running with another application. 
Stalls, for example, caused by true data dependencies or control hazards, rise with inter-application interference. However, not all the stalls grow following the same trend with interference. Therefore, the execution time (cycles) due to stalls needs to be distributed into categories, each one representing a component of the execution time affected in a distinct way by interference.

\subsection{Building an ISC Stack}

A common way of characterizing the execution time of an application, showing where execution cycles are spent, is by building an instruction and stall components (ISC) stack. 

The first decision to characterize performance and build an ISC stack is to decide the point at which the stack will be built. If the ISC stack is built at the issue stage, the instructions component will refer to the issued instructions, and the stalls component will account for issue slots where no instruction has been issued. If we model performance at any other stage (for instance, dispatch or commit), both components should reflect whether instructions progress or not through that particular stage in the pipeline. Notice that a necessary condition to select a given stage is that the target processor provides enough performance counters to build the whole ISC stack. 
That is, the stack height should reach 100\% of the execution cycles to fully characterize the behavior of the application.

\begin{figure} [b]
\centering
\includegraphics[width=0.3\columnwidth]{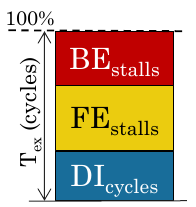}
\caption{
Simple ISC stack gathered at the issue stage.}
\label{fig:pila_procesador_ideal} 
\end{figure}

Let us assume that we are building an ISC stack for the issue stage. After we set the processor stage where performance will be modeled, we can refine the distribution of stalls in different components. As mentioned above, stack components should be chosen according to their behavior when growing due to inter-application interference. In other words, stalls that exhibit similar behavior are grouped into the same component. A simple way is to group the processor stalls according to two main \emph{big} components: the \emph{frontend component} (e.g., instruction cache), whose aim is to feed the processor backend with instructions to be executed, and the processor \emph{backend component} consisting of those functional units (arithmetic operators and caches) that execute the instructions. Figure \ref{fig:pila_procesador_ideal} shows an example of a simple ISC consisting of these three components: Backend stalls ($BE_{stalls}$), Frontend stalls ($FE_{stalls}$), and Dispatched Instructions ($DI_{cycles}$) for a hypothetical application, which we discuss next. The height of each component represents the fraction of the execution time it takes.


These big stack components (also referred to as categories in this work) are, in turn, composed of smaller components. For instance, the Backend category includes both the memory components (e.g., L1 data cache and the main memory) and the core components (e.g., floating-point operators), which are likely not growing at the same pace with inter-application interference. In such a case, a stack might introduce significant error deviations when used by a performance model. Therefore, at first glance, an ISC stack with more components may provide a finer description of how the application behaves, identifying the critical components (if any) that are preventing the application from reaching higher performance. 

Nevertheless, when the ISC stack is to be used by a performance model that estimates how well two applications will co-run in an SMT core, which is the final goal of our work, a higher number of components might not be the best option. First, a higher number of components implies higher overhead to estimate the performance of a combination of applications. Low overhead is a desirable feature because the possible number of combinations quickly grows with the number of cores and applications. In addition, a high number of components can also lead to higher deviations in performance estimation. These deviations strongly depend on the accuracy and \emph{consistency} of the available performance counters in the target processor. But, based on our experience, accumulating small deviations for several components tends to result in a higher overall deviation.
For instance, in our ARM processor, dispatch stalls can be broken down into sub-categories according to distinct hardware events defining the stall cause; however, the sum of all these event counts is unlikely to match the total amount of dispatch stalls exactly.

\subsection{Challenges that Arise with a Real PMU}

So far, we have assumed that any event required to build the ISC stack can be gathered with the PMU deployed in the target machine. In practice, despite current PMUs allowing performance counters to gather hundreds of events, most of them cannot be used effectively for monitoring all the necessary events to build the stack. Instead, a rigorous analysis must be done to overcome PMU limitations.

PMU capabilities strongly depend on the processor architecture.
Moreover, it changes across microprocessor generations within the same architecture, even though these changes are usually smaller.
To focus the discussion, this paper considers an ARM processor.
ARM processors are increasing their market share, not only in embedded systems but also in the server segment. Unlike Intel and IBM POWER processors, performance modeling on ARM processors has received little attention from the research community, mainly because they arrived later at the high-performance market. However, the general insights of our discussion can be extended to other architectures as well.

The first step in building an ISC stack for accurate performance estimates involves analyzing the multitude of events of the target PMU. This analysis aims to identify a pipeline stage where the processor measures both the progression of instructions and the reasons for instruction stalls. Upon identifying such a pipeline stage, the next step is to determine if the events can be distinguished among different stall components, ensuring that the total sum of instructions and stalls accounts for 100\% of the execution cycles. Selecting the stalls may be precise if they originate from a limited number of sources and are captured by well-documented events, but in this case, the stalls arise from numerous sources and are documented ambiguously, complicating the analysis. In such cases, it is crucial to verify that all stalls are accounted for and that each stall is counted only once.

The first limitation we encounter when analyzing the PMU of the experimental ARM ThunderX2 processor and the events it can monitor is that it is only able to capture stalls at the dispatch stage. This means that the ISC stack can only be built at that pipeline stage. Enabling stack building at just one stage is not a particular limitation of ARM processors but rather a common case across different processor manufacturers: Intel processors count stalls at the issue stage, and IBM processors count them at the commit stage. To the best of our knowledge, no commercial processor supports monitoring stalls at multiple pipeline stages. 

The mentioned limitation has important implications. In particular, building the ISC stack at a different stage means not just that stalls are gathered earlier or later in the pipeline but that different types of events will be involved. For instance, the component referred to as \emph{core stalls} \cite{top-down} in the Intel processor is captured at the issue stage and counts the issue slots where instruction can not be issued due to the target functional unit being busy; however, such stall can not be gathered at the dispatch stage in ARM processors because the instruction has not reached yet the instruction queue at the time stalls are gathered.

In summary, although ISC stacks can be built at distinct stages depending on the target processor, both the stage and stall categories must be selected under a thorough examination of the processor's microarchitecture and PMU capabilities.

\section{Facing Challenges in a real PMU: Building an ISC Stack for the ARM ThunderX2}

\subsection{Basic ISC Stack}

\begin{table}[t!]
\centering
\caption{Hardware events gathered in the ARM processor to characterize the performance of applications.}
\resizebox{\columnwidth}{!}{%
\begin{tabular}{|l|l|}
\hline
Counter name        & Explanation                                                                   \\ \hline
CPU\_CYCLES         & Cycles                                                                        \\ \hline
\multirow{2}{*}{STALL\_FRONTEND}     & Cycles on which no operation is dispatched because \cr & there is no operation in the queue    \\ \hline
\multirow{2}{*}{STALL\_BACKEND}      & Cycles on which no operation is dispatched due to \cr & backend resources being unavailable \\ \hline
INST\_RETIRED       & Instruction architecturally executed                                          \\ \hline
INST\_SPEC          & Operation (speculatively) executed                                              \\ \hline
\end{tabular} 
}
\label{tab:performance-counters-arm}
\end{table}

Table~\ref{tab:performance-counters-arm} presents the events monitored at the dispatch stage, which we will use to build the stack, including frontend and backend stalls, according to the description publicly available in the documentation \cite{armv8}. Stalls are attributed to the frontend when the dispatch queue is empty and assigned to the backend when a backend resource (e.g., an ROB entry) is not available, preventing the dispatch of instructions. This means that in a basic ISC stack, a \emph{Frontend category} could be easily computed by dividing the \emph{STALL\_FRONTEND} event by cycles (i.e., \emph{cpu\_cycles} event) to express it as a fraction of the execution time. The \emph{Backend category} could be similarly computed with the \emph{STALL\_BACKEND} event.

The processor PMU does not explicitly count cycles on which instructions are dispatched. In principle, it is safe to assume that in those cycles not accounted as stalls, at least one instruction has been dispatched. This implies that these cycles comprise both those where all dispatch bandwidth is consumed and those where fewer instructions are dispatched. That is, the instructions category in the ISC stack includes cycles with a wide variety of consumed dispatch slots (i.e., ranging from 1 to the dispatch width), so significant error deviations would be introduced if the stack is used for performance estimates. The main reason is that a cycle on which a single instruction is dispatched behaves closer to a stall cycle, while a cycle on which three instructions are dispatched behaves closer to a cycle where four instructions are dispatched (a \emph{full dispatch} cycle). 
Putting them together in a category later used by a performance prediction model will introduce significant noise into the predictions, which is a challenge that we have to deal with since we do not work with an ideal PMU to build the ISC stack.
However, since no event in the ARM ThunderX2 PMU allows for the classification of cycles according to the number of instructions dispatched, we use the PMU to estimate the number of full dispatch cycles required to dispatch a number of instructions equivalent to those dispatched in a period where the number of consumed dispatch slots changes every cycle and refer to these cycles as \emph{full dispatch equivalent} cycles.

The PMU provides two main events referring to instructions flowing through the pipeline (see Table~\ref{tab:performance-counters-arm}): retired instructions (\emph{INST\_RETIRED} event) and executed instructions (\emph{INST\_SPEC} event). From these counters, we opt to approximate the number of dispatched instructions using \emph{INST\_SPEC} as the execution stage is closest to the dispatch stage. 
That is, we only gather \emph{INST\_RETIRED} event to use it for the evaluation of performance, not to build the ISC stack.
Differences will appear since not all the dispatched instructions are executed, as some of them can be squashed before execution (e.g., due to a mispredicted branch). However, due to the high accuracy of current branch predictors and considering that some instructions on the wrong path are executed before the branch instruction is resolved, this event is expected to closely approximate a hypothetical (and non-available) \emph{dispatched instructions} event. From now on, \emph{INST\_SPEC} will be used as an estimate of the number of dispatched instructions. 

According to the previous rationale, the ISC category that represents the fraction of execution cycles taken by full dispatch equivalent cycles (\emph{dispatched instructions cycles} category, or merely \emph{dispatch cycles}) can be estimated by dividing the executed instructions by the dispatch width and the execution time; that is, \textit{INST\_SPEC / (4$\times$ CPU\_CYCLES)}. This formula computes the fraction of cycles that would have been consumed to dispatch an equivalent number of instructions at full dispatch bandwidth.

\subsubsection{Characterization}

In the previous section, we discussed a basic ISC stack with three categories: Dispatched Instructions cycles ($DI_{cycles}$), Frontend stalls ($FE_{stalls}$), and Backend stalls ($BE_{stalls}$).  
This section presents the ISC stacks obtained experimentally with the performance counters of the ARM ThunderX2 processor and SPEC CPU workloads (see Section \ref{sec:experimental_framework} for further details on the experimental platform and workloads).

\begin{figure}[b!]
    \centering  
        \includegraphics[width=0.5\linewidth]{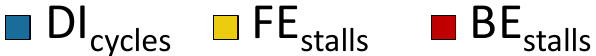}
        \centering
        \includegraphics[width=\linewidth]{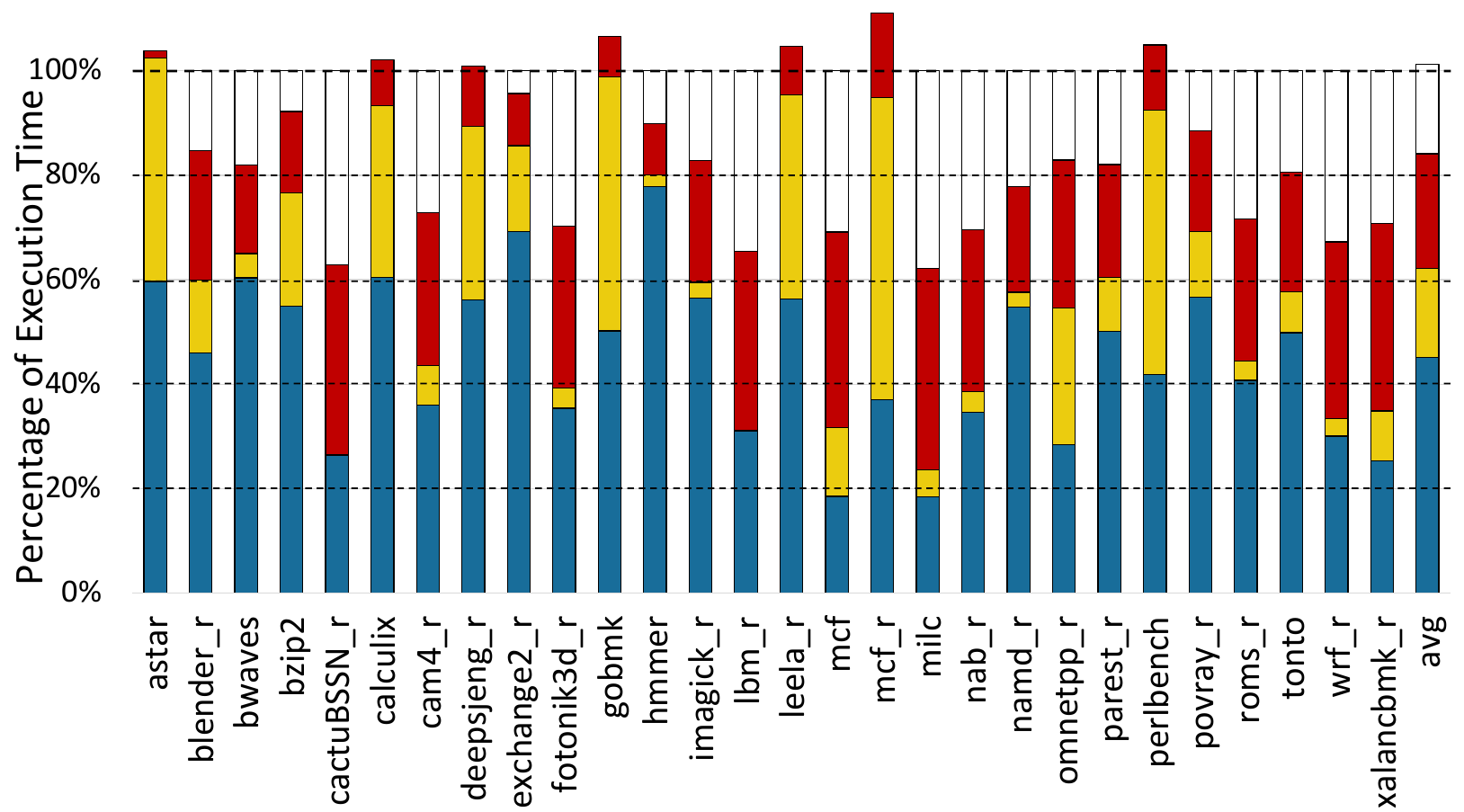}
    \caption{ISC stacks of applications in isolated execution.}
    \label{fig:caracterizacion-versiones}
\end{figure}

Figure \ref{fig:caracterizacion-versiones} presents the ISC stacks for each application in individual execution (i.e., alone in the processor). These stacks have been built with the three categories studied above, gathering the required performance counters at the end of the execution. At first glance, two main cases can be observed: applications whose stack is less than 100\% of the execution cycles and applications whose stacks are over 100\%. The top box of the bar, colored in white, represents the non-captured cycles in those stacks below 100\%.

In principle, one expected that the sum of the three studied categories would be below 100\%. The reason is that the dispatch category is obtained by converting all the cycles where at least one instruction is dispatched in the equivalent number of cycles if four instructions were dispatched. Thus, this is the common case, and a significant set of applications, about 75\% of the benchmarks (21 out of 28), present a performance stack below 100\%.

Surprisingly and counterintuitively, it can be observed that the sum of these three categories in some cases exceeds 100\% of the processor cycles. Just 7 applications suffer this issue, and some of them barely exceed 100\% of the execution cycles. Note, however, that \texttt{mcf\_r} surpasses this threshold by 15\%. The main reason is how the PMU works. The PMU is not designed to implement the ISC stack but to count hardware events. Thus, counters work independently of each other, and thus, multiple events can be accounted for in the same cycle. Intel provides a solution for this issue in some specific cases; for instance, backend events are tested before frontend events so that if both a backend stall and a frontend stall rise at the same cycle, the stall is assigned to the backend category \cite{top-down}. To the best of our knowledge, this is not documented for ARM processors.

Most of these stacks obtained experimentally, in their current way, cannot be used as inputs for a performance model due to some of them falling far from 100\% of the execution cycles. In applications such as \emph{cactuBSSN\_r}, \emph{lbm\_r}, or \emph{milc}, the non-represented (white box) cycles range in between 35\% about 40\% of the execution cycles. This large percentage demands a careful analysis to devise alternative approaches to expand the performance stack up to 100\%. Similarly, stacks over 100\% should be shrunk to fit 100\%.

In summary, according to the represented cycles, two main types of stacks can be distinguished: those capturing less than 100\% of the cycles (case LT100) and those exceeding this value (case GT100). Next, we discuss the proposed approaches to adjust both cases to match the 100\% of the execution cycles.

\subsection{Case LT100} 
\label{sec:lt100}

\begin{figure}[t!]
\centering
\vspace{0.05cm}
\includegraphics[width=\columnwidth, trim=0 10 0 10, clip]{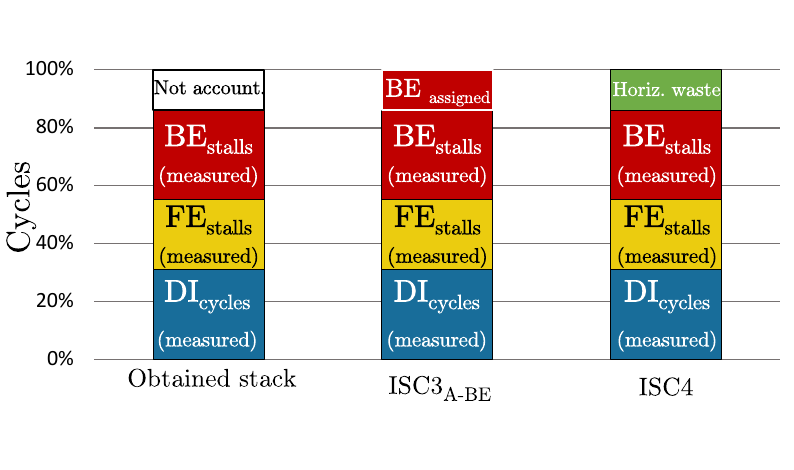}

\caption{Devised approaches to deal with Case LT100.}

\label{fig:fase1_2opciones} 
\end{figure}

This section aims to expand the ISC stack in the LT100 case to the total execution cycles.

The left column (obtained stack) in Figure \ref{fig:fase1_2opciones} represents the measured categories as described in the previous sections. The bar's top box (Not account. label) represents the execution cycles not accounted for in the stack. The remaining columns depict the approaches devised. The number in the approach name indicates the number of categories of the ISC stack after expanding the stack gathered online.  

Two main ways approaches have been devised to expand the stack to 100\%, referred to as $ISC3_{A-BE}$ and $ISC4$. 
The former approach, namely $ISC3_{A-BE}$, assigns the not accounted cycles to the Backend category ($BE_{assigned}$ label). This seems reasonable in most applications as the backend, mainly cache hierarchy and external main memory, are typically the major contributors to stalls. 
This approach is the one used in our previous work \cite{ipdps-24}.
This can be observed in our experimental platform in most stacks falling within the LT100 case (see Figure~\ref{fig:caracterizacion-versiones}). 

The latter approach, $ISC4$, adds a new category, which is referred to as \emph{Horizontal waste} (Horiz. waste label). This category counts the cycles on which between one and three instructions are dispatched. The reason to add this new category is that the \emph{horizontal waste cycles} could not necessarily behave as the Backend stalls cycles category. Next, we discuss this claim. 
Typically, this happens when the ROB is full and a long memory latency instruction blocks the ROB's head. Thus, an important number of stalls can rise when this happens, which can significantly grow with the number of co-runners. In contrast, horizontal waste refers to partial stalls, as some instructions are still dispatched within these cycles and are usually triggered by intra-core interference. 

Differentiating both cases is fundamental when the ISC stack is used as input for a performance model. In other words, from a theoretical perspective, and as experimental results will show, it is expected that $ISC4$ provides better accuracy when used by a performance model than $ISC3_{A-BE}$.

\subsection{Case GT100}
\label{sec:gt100} 
As previously discussed, the sum of the categories can exceed 100\% of cycles. This situation occurs because some hardware events partially overlap, and multiple stall causes are recorded in a single cycle. We resolve this excess of cycles by subtracting the exceeding fraction from some of the categories of the ISC stack so that it fits the 100\% of execution cycles. The question is from which categories should they be subtracted to fit 100\%.

\begin{figure}[t!]
\centering
\includegraphics[width=\columnwidth, trim=0 0 0 10.5, clip]{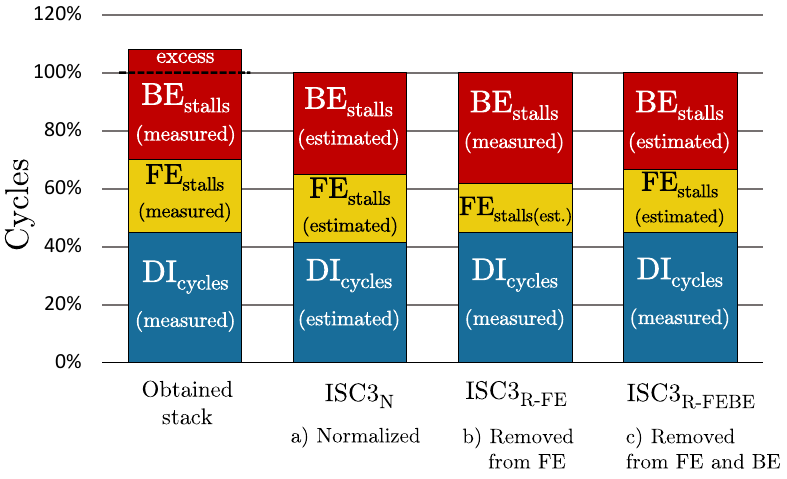} 
\caption{Three main ways to deal with Case GT100.}
\label{fig:fase2_3opciones} 
\vspace{-0.3cm}
\end{figure}

Figure \ref{fig:fase2_3opciones} illustrates this scenario and the devised approaches. The first bar, labeled as `Obtained stack,' represents the ISC stack gathered experimentally. It considers the categories $BE_{stalls}$, $FE_{stalls}$, and $DI_{cycles}$, and shows the hypothetical `excess' of cycles. 
Again, performance counters do not give us any hint to accurately assign these \emph {exceeding cycles}. To deal with this situation, we devise three main approaches: $ISC3_{N}$, $ISC3_{R-FE}$, and $ISC3_{R-FEBE}$, discussed below.

The $ISC3_{N}$ approach, used in \cite{ipdps-24}, assumes that the three components of the `Obtained stack' contribute to the overlapping cycles proportionally to their weight in the stack. This approach handles the excess of cycles by subtracting the corresponding fraction of the exceeding cycles from each component, i.e., all the categories are normalized so that the ISC stack sums 100\%. Thus, for this approach, all the normalized categories are labeled as \emph{estimated} in the figure.

As mentioned above, overlapping occurs because multiple counters are increased at the same time; that is, the stall hardware events may account multiple times for the same stall in a single cycle, assigning it to the Backend and Frontend categories. Thus, it seems a better approach will assign overlapping hardware to a stall category. 
In this way, the $ISC3_{R-FE}$ approach considers that the exceeding cycles are mainly caused by the Frontend stalls ($FE_{stalls}$) component. This assumption seems reasonable for the target machine, where this category seems excessive compared to other machines like Intel Xeon processors\footnote{Note that we do not consider a possible converse $ISC3_{R-BE}$ approach because it will grow the Frontend stalls category even more.}.
Thus, this approach subtracts all the exceeding cycles (excess) from this component. 
Therefore, only the value $FE_{stalls}$ is considered as estimated since the rest of the values remain unchanged with respect to the obtained stack.

In contrast, the $ISC3_{R-FEBE}$ approach assumes that the exceeding cycles can be equally due to both stall categories ($BE_{stalls}$, and $FE_{stalls}$). Thus, to fit the stack to 100\%, this scheme subtracts the corresponding fraction of the exceeding cycles from both components. In this case, both categories are considered estimated.

\section{SYNPA Family of SMT Scheduling Policies}

\subsection{SYNPA Family Members} \label{sec:synpa_versions}

In the previous section, we have elaborated on different ways to build the ISC stacks according to two main cases referred to as LT100 and GT100. SYNPA relies on using a specific way for each case. 
In \cite{ipdps-24}, we evaluated the combination presented in the first row of Table~\ref{tab:SYNPA-versions-ISC}, that is, three categories for the case LT100 and normalized categories for the case GT100. 
This variant is referred to in this paper as $SYNPA3_{N}$. 

To homogenize the terminology, the members of the family or SYNPA variants are referred to as $SYNPAX_{ZZZ}$ where $X$ can be 3 or 4 depending on the categories composing the ISC stack for the case LT100, and $ZZZ$ refers to the ISC stack method used for the case GT100.

\begin{table}[t!]
\caption{SYNPA variants and the applied  ISC stack method.}
\centering
\begin{tabular}{|l||l|l|}
\hline
\multicolumn{1}{|c||}{\multirow{2}{*}{Algorithm}}  & \multicolumn{2}{c|}{Measured}                                            \\ \cline{2-3}
 & \multicolumn{1}{c|}{$ < 100\%~cycles$}     & \multicolumn{1}{c|}{$> 100\%~cycles$} \\ \hline \hline
$SYNPA3_{N} $          & $ISC3_{A-BE}$  & $ISC3_{N}$       \\ \hline
$SYNPA4_{N}$           & $ISC4$         & $ISC3_{N}$       \\ \hline 
$SYNPA4_{R-FE}$        & $ISC4$         & $ISC3_{R-FE}$    \\ \hline 
$SYNPA4_{R-FEBE}$      & $ISC4$         & $ISC3_{R-FEBE}$  \\ \hline
\end{tabular}
\label{tab:SYNPA-versions-ISC}
\end{table}

This paper presents three new variants to the SYNPA family, all of them using four categories to build the stack in the case LT100. The new members of the family are named $SYNPA4_{N}$, $SYNPA4_{R-FE}$, and $SYNPA4_{R-FEBE}$.
Table~\ref{tab:SYNPA-versions-ISC} summarizes the methods used to build the stacks for these new variants.

\subsection{Estimating Performance Using Linear Regression}

SYNPA uses a small set of linear regression models, one per category of the ISC stack, to estimate the performance degradation of each application with each possible co-runner.
Using an independent equation per category presents a first insight into which category (i.e., processor components) is limiting the performance of each individual application.

The linear regression equation used is based on Equation \ref{eq:model-regression}. The output  $C^{smt}_{i,j}$ of this equation represents the value of category $C$ for application $i$ being executed in the same core with application $j$ in SMT mode. 
$C^{st}_{i}$ and $C^{st}_{j}$ represent the value of category $C$ for application $i$ and $j$, respectively, when each of these applications is executed alone in the system in Single-Threaded (ST) mode. The model coefficients, $\alpha_{C}$, $\beta_{C}$, $\gamma_{C}$, and $\rho_{C}$, vary according to the category; and are independent of the running application.

\begin{equation} \label{eq:model-regression} 
{C}^{smt}_{i,j} = \alpha_{C} + \beta_{C}\cdot C^{st}_{i} + \gamma_{C}\cdot C^{st}_{j} + \rho_{C}\cdot C^{st}_{i}\cdot C^{st}_{j}
\end{equation}

In summary, the equation is composed of three main components: the value of $C$ in ST mode ($C^{st}_{i}$), the value of $C$ of the co-runner $j$ in ST mode ($C^{st}_{j}$), and the product of these components ($C^{st}_{i}\cdot C^{st}_{j}$). These components are multiplied by a coefficient, $\beta_{C}$, $\gamma_{C}$, and $\rho_{C}$, respectively, which balance the effect of each component for each category. The coefficient $\alpha_{C}$ is the independent term that aims to reduce the inaccuracies of the model, so minimizing possible deviations. 

\begin{table}[t!]
\caption{ 
Model coefficients for each component of $SYNPA3_N$ and $SYNPA4_N$.
}

\centering

 \begin{tabular}{|c|r|r|r|r|}
  \multicolumn{5}{c}{ \multirow{2}{*}{$SYNPA3_N$} }           \\ 
\multicolumn{5}{c}{} \\ \hline
    Category & \multicolumn{1}{c|}{$\alpha$} & \multicolumn{1}{c|}{$\beta$} & \multicolumn{1}{c|}{$\gamma$} & \multicolumn{1}{c|}{$\rho$}  \\ \hline
    Dispatch       & 0.0072  & 0.9060  & 0.0044  & 0.0314 \\ \hline 
    Frontend       & 0.2376  & 1.4111  & 0       & 0      \\ \hline 
    Backend        & 0.2069  & 0.3431  & 1.4391  & 0      \\ \hline 
    
\multicolumn{5}{c}{ \multirow{2}{*}{$SYNPA4_N$} }           \\ 
\multicolumn{5}{c}{} \\ \hline
 Category & \multicolumn{1}{c|}{$\alpha$} & \multicolumn{1}{c|}{$\beta$} & \multicolumn{1}{c|}{$\gamma$} & \multicolumn{1}{c|}{$\rho$}  \\ \hline
    Dispatch         &  0.0070 & 0.9090 & 0.0021  & 0.0312 \\ \hline 
    Frontend         &  0.2358 & 1.4147 & 0       & 0      \\ \hline 
    Backend          &  0      & 0.2401 & 1.0654  & 0      \\ \hline 
    Horiz. Waste         &  0.2899 & 0.3306 & 1.6111  & 0      \\ \hline 
    \end{tabular}%
    
\label{tab:coef-linear-regression}
\end{table}

As this paper proposes and evaluates four main SYNPA policies, and the categories vary with each policy, the coefficients are obtained for each category and SYNPA version.
For the sake of clarity and illustrative purposes, Table \ref{tab:coef-linear-regression} shows the coefficients of $SYNPA3_{N}$ and $SYNPA4_{N}$ that have been obtained following the methodology discussed in Section \ref{sec:training}. 
The Dispatch and Frontend categories are calculated similarly in both versions. As a consequence, there is a similarity between the values of their coefficients and their corresponding Mean Square Error (MSE) obtained to evaluate the accuracy of the model.
The MSE values are around 0.0021 and 0.00703 for the Dispatch and Frontend categories, respectively.

Regarding the coefficients, in the Dispatch category, the $\beta$ coefficient shows that the execution of a given application progresses slowly in SMT mode compared to in isolation (ST mode) since, in SMT execution, the interference grows, which translates into an increase in the number of stalls. In addition, the $\rho$ coefficient indicates that the interference between the threads affects the dispatch rate of the application in SMT mode.
With respect to the Frontend category, SMT execution increases more than ST, and this effect does not depend on the co-runner. 
Notice that this category needs a non-negligible value of $\alpha$.
That is, the value in SMT mode for this category is difficult to predict, mainly when applications constrained by the frontend resources are executed together. 
The Backend category of $SYNPA3_{N}$ splits into two main categories, Backend and Horizontal waste in  $SYNPA4_{N}$, which makes the coefficients of both Backend categories differ.
These differences mainly appear in the  $\alpha$ and $\gamma$ coefficients. At first glance, the $\alpha$ parameter for the Horizontal waste category in $SYNPA4_{N}$ could seem a bit high; the main reason is that this value significantly grows with intra-core thread interference in SMT mode. On the other hand, the value of the $\gamma$ coefficient shows that the behavior of the co-runner is notably important in order to estimate the Backend category. Notice that by separating the Backend and Horizontal waste categories, the values of the coefficients of the new Backend category better fit the application's behavior. For instance, $\alpha$ goes from 0.2069 to 0.

In terms of MSE, $SYNPA3_{N}$ obtains 0.1583 for the Backend category, and $SYNPA4_{N}$ obtains 0.0277 for the same category and 0.0874 for the Horizontal waste. 
These MSE values show that including the backend stalls and horizontal waste within a single category makes the behavior of the resulting category more difficult to estimate, leading to significant inaccuracies.


\subsection{SYNPA Methodology: Steps Followed} \label{sec:steps-synpa}

SYNPA policies apply once per quantum during the workload execution. The quantum length has been set to 100ms in the experiments. After each quantum expires, the performance events in Table \ref{tab:performance-counters-arm} are gathered to compute the categories and build the performance stack of each application.

\begin{figure}[b!]
\centering
\includegraphics[width=\columnwidth]{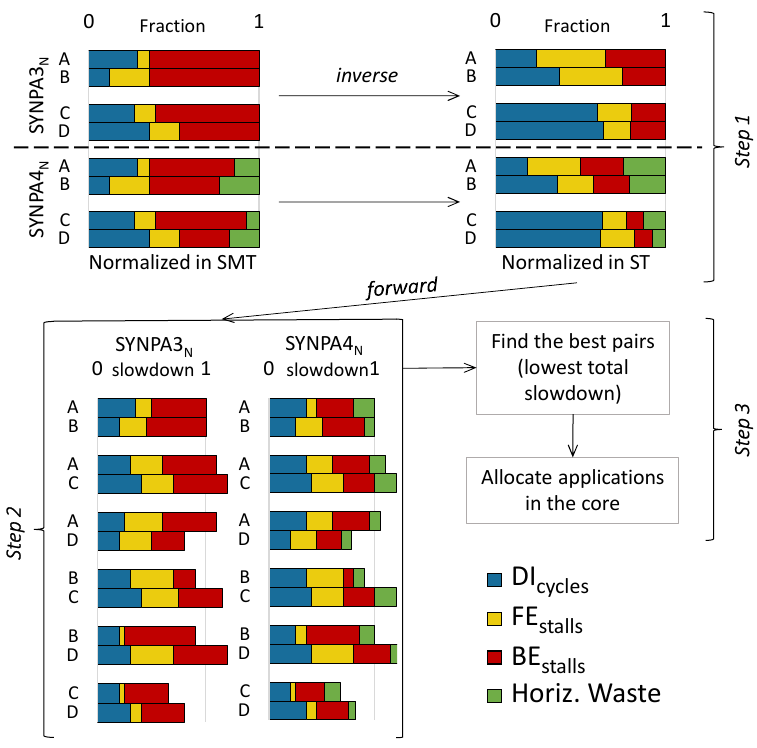}
\caption{General overview of the steps in $SYNPA3_{N}$ and $SYNPA4_{N}$.}
\label{fig:general-steps} 
\end{figure}

Once the performance stack is built for a specific SYNPA policy, we must apply the performance prediction model to select the best pairs of applications for each core. This process consists of three main steps: i) applying the inverse model to obtain the ISC stack in ST execution, ii) applying the forward model to estimate the slowdowns in SMT execution, and iii) choosing and allocating the best pairs in cores.  Figure \ref{fig:general-steps} depicts these steps for $SYNPA3_{N}$ and $SYNPA4_{N}$ discussed below.

\begin{description}[wide=0pt] 

\item[Step 1.]~\textbf{Obtaining the ISC stack in ST: Inverse Model}

The regression model discussed above estimates the slowdown of every possible pair when running together using the values of the categories of each application running alone in the system (i.e., in ST mode) as input.
However, these inputs cannot be collected during SMT execution. 
This problem can be solved by applying the regression model in an \emph{inverted manner} ~\cite{symbiotic-josue} to the obtained categories in SMT execution to build the ISC stack each application would have had running alone in the system (i.e., the ISC stack in ST mode). Once the values of the ISC for ST execution are obtained, they are normalized so that the stack fits 1. 
This step is depicted in the top-right side of Figure \ref{fig:general-steps} for $SYNPA3_{N}$ and $SYNPA4_{N}$.

\item[Step 2.]~\textbf{Obtaining the ISC stack in SMT: Forward model.}

The forward model is applied once the values of the categories for each application in ST mode are estimated. This model, corresponding to Equation \ref{eq:model-regression}, estimates the performance degradation (slowdown) for a pair of applications. The model equation needs to be applied twice for each pair of applications as it is applied once for each application of the pair to get its individual impact on performance. 
These equations capture the effect in each category that the applications have on each other, estimating their synergy if they were executed on the same core. 
As it can be seen in the bottom-left side of Figure \ref{fig:general-steps} (Step 2), both SYNPA versions gather the values of the slowdown of each possible pair of applications. Note that the application could not show the same behavior in both SYNPA versions since the category values differ.

\item[Step 3.]~\textbf{Selecting and allocating the best pairs.} 

Finally, the Blossom Algorithm~\cite{blossomalgorithm} is applied to obtain the best combination of pairs of applications, i.e., the combination of pairs with the lowest performance degradation. This algorithm considers all the possibilities and selects the optimal choice with minimum overhead, even if the number of applications increases. 
As depicted in Step 3 of Figure \ref{fig:general-steps}, once the best pairs are found, the applications are allocated to their corresponding core to execute the most synergistic combination in the next quantum. 
\end{description}

\subsection{Model Building} \label{sec:training}


A model needs to be built for each policy, that is, for each SYNPA variant. In Figure~\ref{fig:caracterizacion-versiones}, we characterized 28 applications. The models were built using 22 out of these 28 applications. The remaining 6 applications (i.e., \texttt{imagick\_r, parest\_r, leela\_r, wrf\_r, cam4\_r}, and \texttt{exchange\_r}) were used to assess their accuracy thus they were not considered to train the models.

To build each model, applications were run in isolation, and then, we created a profile with the value of the different categories and the number of committed instructions for each quantum. 
Next, we run all the possible pairs of these applications in SMT mode and collect the same data. As execution progresses slower in SMT mode, the number of committed instructions allows us to map the category values of an application when it runs in isolation to the corresponding values when the application runs in SMT mode with another application. We do not need to use all the data gathered from performance counters across all the quanta but to capture the distinct behaviors of each application. Thus, to save time building the model, a random subset of the execution quanta was selected. The coefficients were calculated, minimizing the model error (MSE) as much as possible.

In this work, the model coefficients were obtained using applications from the SPEC CPU suite. 
This means that the performance of applications exhibiting a similar range of behaviors will be accurately predicted. 
Hence, the model can be used to select the most synergistic pairs of applications as long as the behaviors of the running applications do not noticeably deviate from those of the applications used to train them. Nevertheless, if the model needs to be applied to applications showing significantly different behaviors, its training can be extended to encompass these new dynamics.

\section{ Experimental Framework} \label{sec:experimental_framework}

\subsection{Platform}

Experimental results have been obtained in a Cavium ThunderX2 CN9975 processor~\cite{ThunderX2} equipped with 64GB DRAM main memory.
The processor is based on the Vulcan microarchitecture~\cite{vulcan}, implements the ARMv8.1A instruction set~\cite{armv8}, and has a 28MB shared LLC. This processor deploys 28 SMT cores.
The system runs a CentOS Linux 7 (AltArch) distribution~\cite{centos-mirror} with kernel 4.18. 

The devised policies have been implemented in a custom framework developed in-house to carry out the performance evaluation.
The framework helps execute workloads and gather the dynamic values of the target performance counters during execution time. It uses the perf tool~\cite{perf-tool} to configure and read performance counters along the execution. To allocate applications to cores, the framework employs the sched\_setaffinity system call~\cite{sched-setaffinity}. 
We also execute these workloads under the Linux OS scheduler (Completely Fair Scheduler or CFS) to compare its performance with those of our policies.
We execute all the experiments within the same framework for a fair comparison with the Linux scheduling policy.

\subsection{Evaluation Methodology}

To evaluate the devised policies, we have used a wide set of 35 workloads, each one consisting of 8 individual applications randomly taken from 24 (18 of them were also used to train the model, and the remaining 6 were reserved for model assessment). 
All the applications are from the SPEC CPU2006~\cite{spec2006} and CPU2017~\cite{spec2017} suites. 
The main aim is to cover a wide range of scenarios. 

As a first step to define the workloads, 28 benchmarks were characterized when running alone. 
Results were presented in Figure \ref{fig:caracterizacion-versiones}. 
A wide range of behaviors can be observed across the studied benchmarks.
Applications can be classified according to their predominant category (biggest value) exhibited during isolated execution. 
Three groups are defined: Frontend-Bound, Backend-Bound, and Others. The Frontend-Bound group includes those applications with a fraction of the Frontend category higher than 35\%, the Backend-Bound group comprises those that have a fraction of the Backend category bigger than 65\%, and Others corresponds to the applications that do not belong to either of the other two groups.  

Three types of workloads are created with these groups of applications: Backend-intensive, Frontend-intensive, and Mixed. To compose each workload, applications for each group have been randomly selected according to the following rules: i) \emph{Backend-intensive workloads,} composed of 5 or 6 Backend-Bound applications and the rest of the applications from the Others group; ii) \emph{Frontend-intensive workloads}, composed of 5 or 6 Frontend-Bound applications and the rest of the applications from the Others group, and iii) \emph{Mixed workloads}, consisting of 4 Backend-Bound applications and 4 Frontend-Bound applications.
We built and evaluated a total amount of 35 workloads: 15 Backend-intensive (\texttt{be0-be14}), 5 Frontend-intensive (\texttt{fe0-fe4}), and 15 Mixed (\texttt{fb0-fb14}).

The methodology followed to gather the statistics for the studied workloads is explained next.
First, each application was run alone for 60 seconds, and the number of instructions committed during this period was gathered. This number was used as the target number of instructions to execute for that application.
During workload execution, the performance counters of each individual application are collected at each (100 ms) interval.
The workload is kept running until the slowest application reaches its target number of instructions.
When an application finishes earlier, then it is relaunched. In this way, the workload runs with a constant number of applications during the whole execution.

The execution of each workload was repeated at least 10 times to obtain reliable values. Then, the coefficient of variation ($\sigma/\mu$) of the execution times was computed and, after that, those execution times over $\mu\pm0.05\times \sigma/\mu$ were discarded, and non-discarded values averaged.

\section{Performance Evaluation} \label{sec:evaluation_versions_synpa}

\subsection{$SYNPA3_{N}$ versus $SYNPA4_{N}$}
This section compares the performance of $SYNPA3_N$ and $SYNPA4_N$ with the aim of identifying the best baseline stack. More precisely, to find out how many categories (3 or 4) help SYNPA to achieve the best performance. The analysis considers the speedup achieved by SYNPA versions over Linux in both Turnaround Time (TT) and the geometric mean of Instructions Per Cycle (IPC).

\begin{figure}[t!]
    \centering   
    \includegraphics[width=0.4\columnwidth]{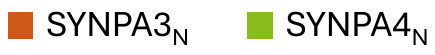} 
    \hfill    
     \subfloat[Turnaround Time (TT) ]{
        \centering
        \includegraphics[width=\columnwidth, trim = 0 70 0 68, clip]{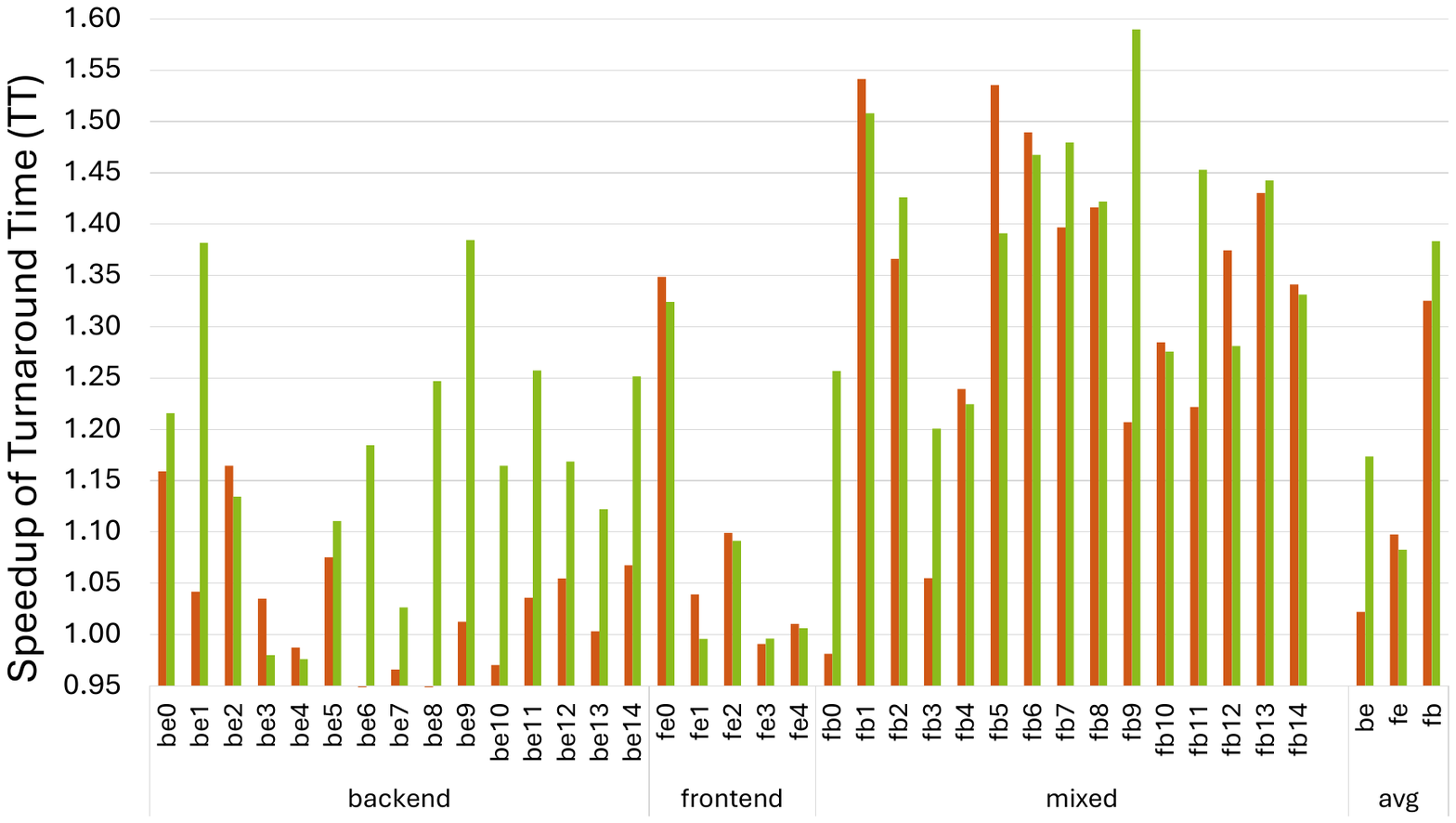}
        \label{fig:TT_fase1_2opciones}
    }
    \hfill
    \subfloat[Instructions Per Cycle (IPC)]{
        \centering
        \includegraphics[width=\columnwidth, trim = 0 70 0 70, clip]{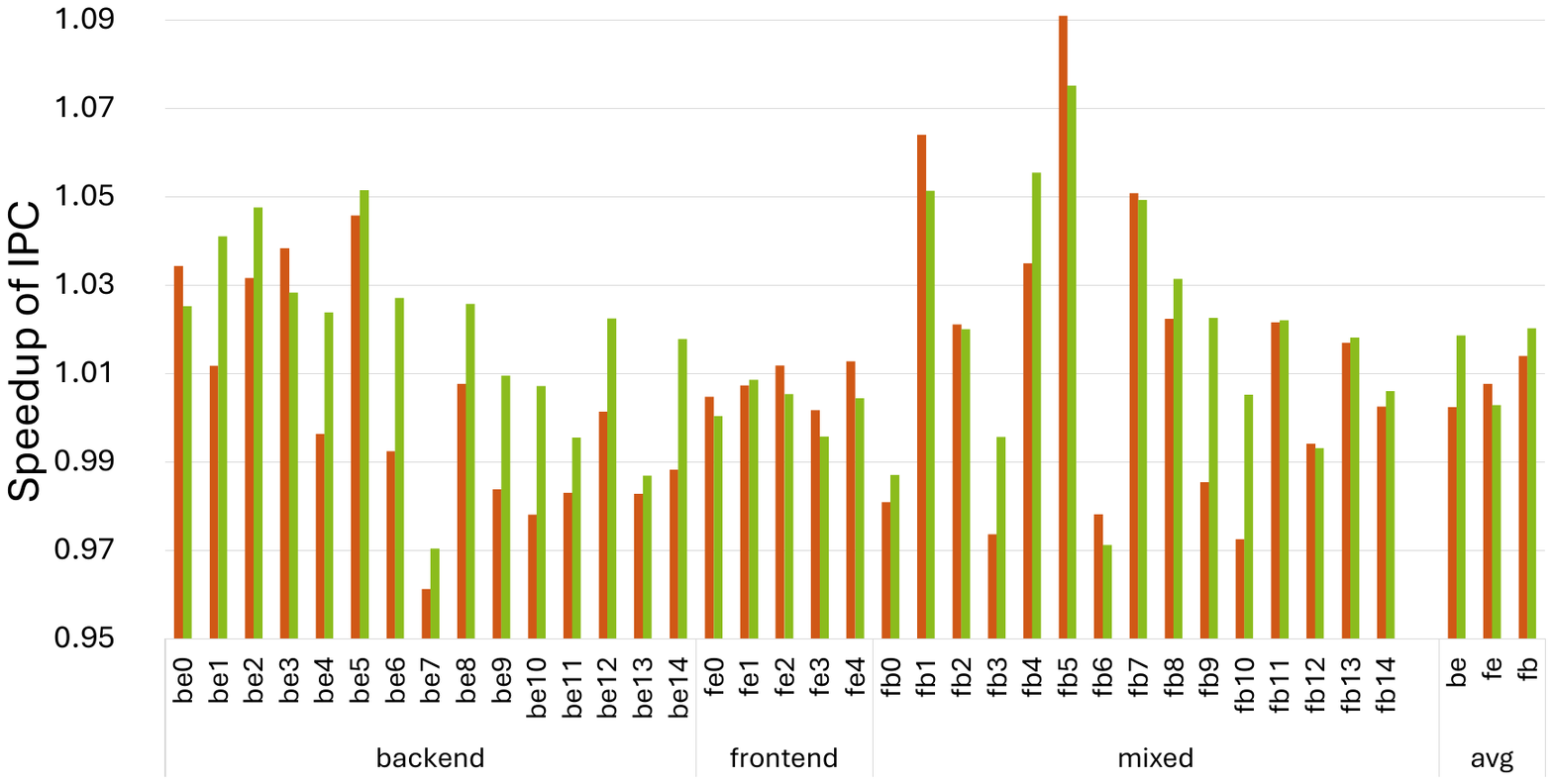}
        \label{fig:IPC_fase1_2opciones}
    }
    \caption{Speedups of $SYNPA3_N$ and $SYNPA4_N$ with respect to Linux.}
    \label{fig:Results_fase1_2opciones}
\end{figure}

Figure \ref{fig:TT_fase1_2opciones} presents the speedups of the TT. As observed, SYNPA achieves substantial performance improvements over Linux, with an average speedup of 38\% for Mixed (fb) workloads, which represent the most common scenario. A comparison between $SYNPA3_N$ and $SYNPA4_N$ shows notable differences in a subset of workloads.
Specifically, in 13 out of the 35 workloads, the performance improvements differ more than 10\% between variants. 
These differences are incredibly high in some workloads; for instance, $SYNPA4_N$ and $SYNPA3_N$ outperform Linux in \texttt{be1} over 38\% and 4\%, respectively,
and in \texttt{fb9}, $SYNPA4_N$ almost triples the performance difference of $SYNPA3_N$.

Figure \ref{fig:IPC_fase1_2opciones} compares the performance improvement of the SYNPA variants regarding IPC. 
On average, both SYNPA versions slightly outperform Linux; however, in certain workloads, performance gains reach up to 9\% and 7.3\% in \texttt{fb5} for $SYNPA3_N$ and $SYNPA4_N$, respectively. 
Notice that in some workloads (e.g., in \texttt{fb3}), Linux achieves slightly better IPC than SYNPA. However, this improvement comes at the cost of TT performance.
In other words, in some workloads, Linux improves IPC by helping the applications with the higher IPC at the expense of the performance of applications with lower IPC, which extends their execution and penalizes the TT.

\begin{figure}[t!]
\centering
\includegraphics[width=\columnwidth]{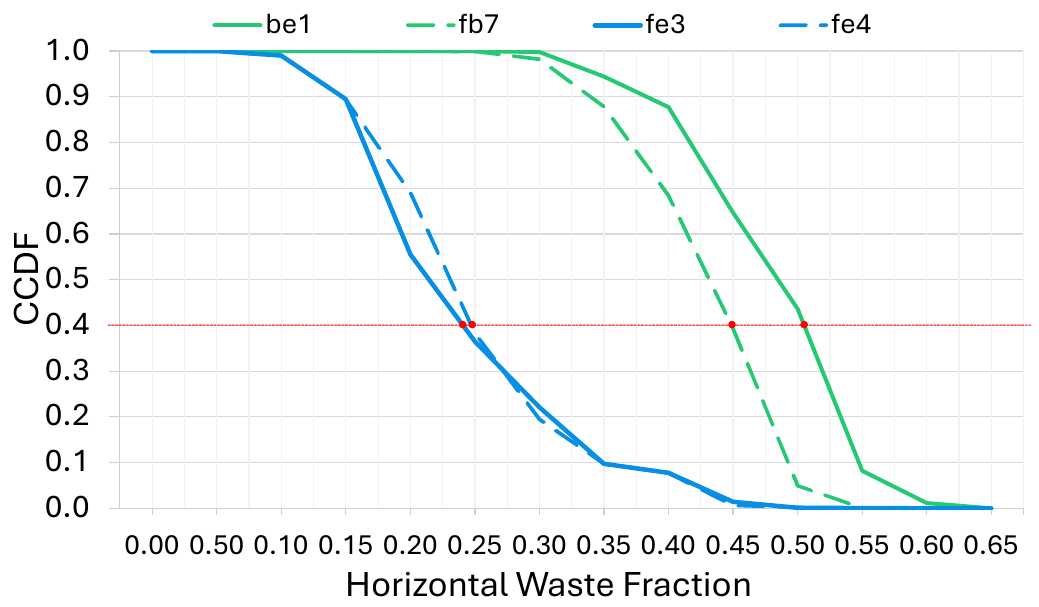}
\caption{Complementary Cumulative Distribution Function (CCDF) of the horizontal waste fraction.} 
\label{fig:cdf-hw-synpa3} 
\end{figure}

The simultaneous analysis of Figures \ref{fig:caracterizacion-versiones} and \ref{fig:TT_fase1_2opciones} reveals that workloads with a higher fraction of horizontal waste in their individual execution benefit the most from splitting the Horizontal Waste and Backend categories. For instance, workloads \texttt{be8} and \texttt{be9}, where applications exhibit a horizontal waste fraction of approximately 0.32, achieve significant TT speedup with $SYNPA4_N$, whereas $SYNPA3_N$ obtains negligible improvements.
On the contrary, workloads \texttt{fb14} and \texttt{fb13} obtain a similar speedup with both SYNPA variants, as their applications do not show horizontal waste. In these cases, treating this category separately does not introduce a noticeable performance difference.
To provide further insights into why huge differences arise, especially in TT, we analyzed how the horizontal waste (i.e., not accounted cycles) is managed in the LT100 case in both SYNPA versions since they only differ in this case (see Table \ref{tab:SYNPA-versions-ISC}).
Recall that $SYNPA3_N$ assigns the horizontal waste to the Backend category, while $SYNPA4_N$ includes a distinct category for it (see Section \ref{sec:lt100}). The main reason behind this independent category is that, intuitively, the \emph{horizontal waste} does not grow as fast as the backend category with interference; thus, it makes sense to consider it as a new category.

Figure \ref{fig:cdf-hw-synpa3} shows the Complementary Cumulative Distribution Function (CCDF) of the horizontal waste fraction for four workloads. Each curve represents the probability (Y-axis) that the horizontal waste fraction exceeds a given value (X-axis), computed as the sum of horizontal waste across all the applications in the workload. 
For instance, workloads \texttt{fe4} and \texttt{be1} have a horizontal waste fraction greater than 0.25 and 0.5 approx. for 40\% (0.4 Y-axis) of their execution time. 
For illustrative purposes, two of the workloads exhibit a high horizontal waste (\texttt{be1} and \texttt{fb7}) across their execution time, and the other two present roughly half this waste (\texttt{fe3} and \texttt{fe4}).


These results prove that the high differences in performance shown in Figure \ref{fig:TT_fase1_2opciones} for some workloads mainly appear with high horizontal waste values. 
For instance, the workloads \texttt{be1} and \texttt{fb7}, where $SYNPA4_{N}$ obtains the biggest improvement over $SYNPA3_N$, are those presenting the highest horizontal waste.

In summary, $SYNPA4_N$ outperforms $SYNPA3_N$ in both TT and IPC, with their only difference being the handling of horizontal waste cycles in the LT100 case. The experimental results demonstrate that $SYNPA3_N$ suffers more performance degradation when the fraction of horizontal waste in the ISC stack is significantly high. Therefore, from the remainder of this work, the evaluation will focus on the four-category stack, that is, in the variants of the $SYNPA4$ policy. 

\subsection{SYNPA4 variants}

\begin{figure}[t!]
    \centering
        \centering
        \includegraphics[width=0.7\columnwidth]{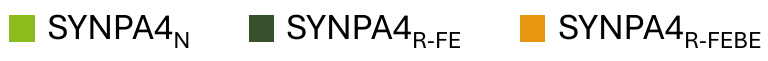}
    \hfill
    \subfloat[Turnaround Time (TT) ]{
        \centering
        \includegraphics[width=\columnwidth, trim = 0 70 0 68, clip]{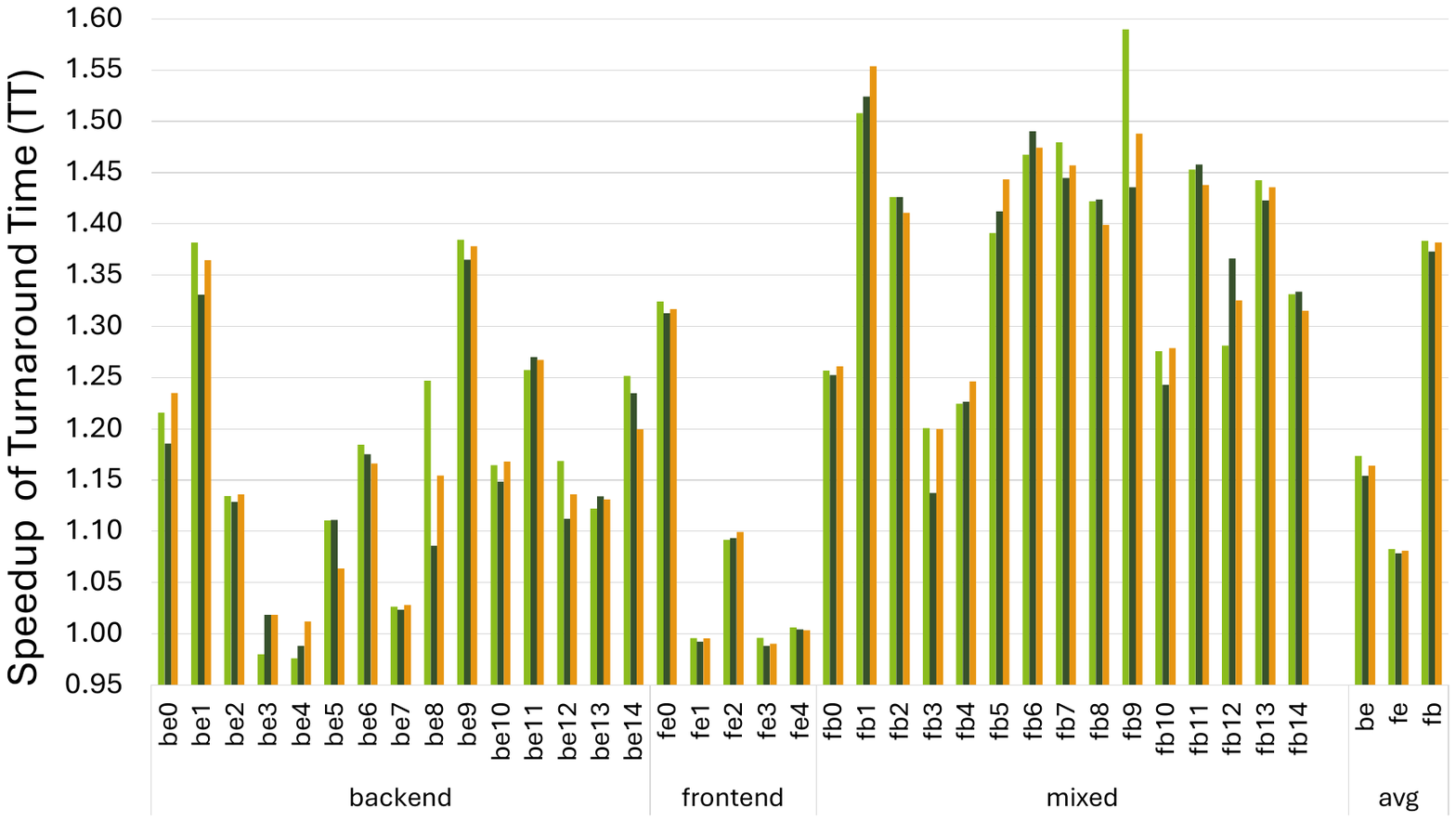}
        \label{fig:TT_fase2_3opciones}
    }
    \hfill
    \subfloat[Instructions Per Cycle (IPC)]{
        \centering
        \includegraphics[width=\columnwidth, trim = 0 70 0 70, clip]{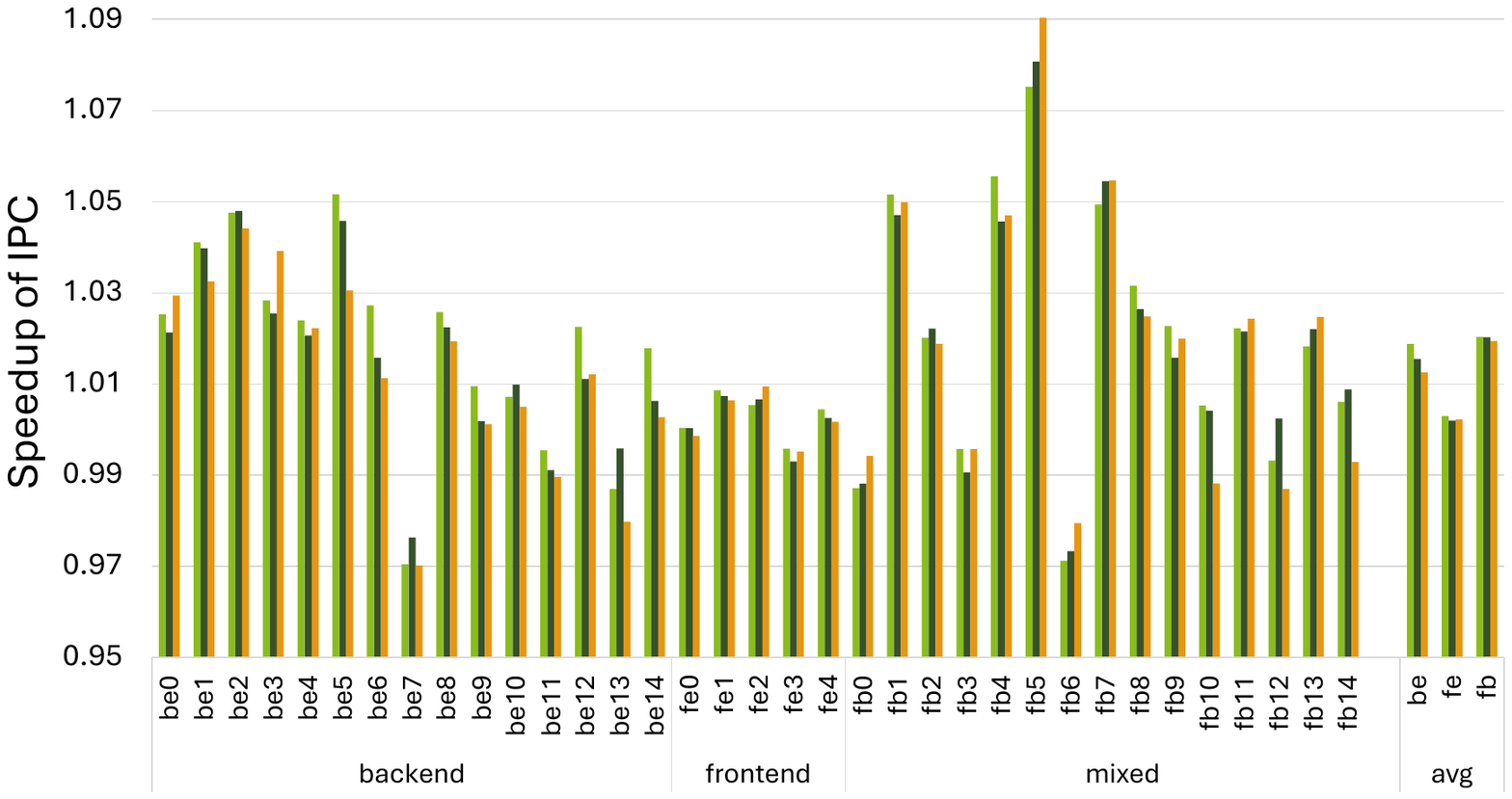}
        \label{fig:IPC_fase2_3opciones}
    }
    \caption{Speedups of SYNPA4 variants.}
    \label{fig:Results_fase2_3opciones}
\end{figure}

This section compares the results of the three variants of $SYNPA4$. 
Remember that all of them use the same ISC stack (ISC4) for those applications having an LT100 ISC stack, but they differ in the ISC stacks for the GT100 case (see Table \ref{tab:SYNPA-versions-ISC}). Therefore, the main aim of this analysis is to identify the best-devised way to handle GT100 ISC stacks. 

Figure \ref{fig:Results_fase2_3opciones} presents the results of the speedup of TT and IPC for the three devised variants. It can be observed that, on average, the three variants slightly differ in performance for both metrics. In some cases, like in \texttt{fb9}, $SYNPA4_{N}$ outperforms the others but drops performance in other workloads like \texttt{fb12}. Due to the minor differences among policies and the strong consistency across the presented results for the studied workloads, we can conclude that none of the approaches studied to handle the GT100 case are better than the others. 
However, as we aim to choose one of them to be implemented in the target policy, we opt for $SYNPA4_{R-FEBE}$, which, on average, is more constant in its performance (high speedups) since analyzing with more detail the speedups always improve Linux in TT or are only slightly worse in some workloads in terms of IPC. In fact, on average, $SYNPA4_N$ may seem to be the best option but is not capable of outperforming Linux in more workloads than $SYNPA4_{R-FEBE}.$

\subsection{SYNPA comparison with other approaches} \label{sec:eval_resul}
In this section, we compare the performance of $SYNPA4_{R-FEBE}$ policy, previously identified as the best approach among the SYNPA variants, against a T2C allocation policy closely resembling Hy-Sched \cite{cal-marta} but adapted to ARM processor, which is a state-of-the-art T2C allocation policy based on heuristics. To make this paper self-contained, we will briefly describe Hy-Sched.

\subsubsection{The Hy-Sched Approach}
The Hy-Sched policy \cite{cal-marta} is based on four main categories: Retiring, Bad Speculation, Frontend-Bound, and Backend-Bound. The two latter categories resemble the frontend and backend stall components in our ISC stacks. The Bad Speculation category refers to cycles on which instructions were issued, but they were later canceled, e.g., due to a mispredicted branch. Finally, the Retiring category refers to cycles on which instructions are successfully committed.
Each individual application of the running workload is dynamically classified according to the category that obtained the highest value (i.e., the largest fraction of its execution) during the last quanta.
Once the four categories have been estimated, Hy-Sched uses a simple heuristic that relies on the fact that pairing two applications from different categories is a sufficient condition to reduce the inter-application interference and improve performance. 
Based on this assumption, the first Hy-Sched option is to pair each application of a workload with an application of a different category. When it is not possible to guarantee that target, the applications are allocated in cores by balancing their IPC (e.g., the application with the highest and lowest IPC are paired).

As Hy-Sched was originally developed for Intel processors, we need to adapt it to the ARM ThunderX2 PMU. The Retiring category is computed in the same way as the PMUs of both architectures can measure the number of retired instructions. To calculate the Frontend-Bound and Backend-Bound categories, we use the frontend and backend stall events, respectively, from the ARM processor. These events resemble the original events used in the Hy-Sched proposal but measure the stalls at dispatch rather than at issue, as done in the Intel PMU. Finally, the Bad Speculation category is computed by subtracting the retired instructions from the speculatively executed instructions.

\subsubsection{Evaluation}
\begin{figure}[t!]
    \centering
    \includegraphics[width=0.45\linewidth]{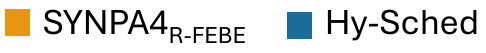}
    \hfill
    \subfloat[Turnaround Time (TT) ]{
        \centering
        \includegraphics[width=\columnwidth, trim = 0 70 0 68, clip]{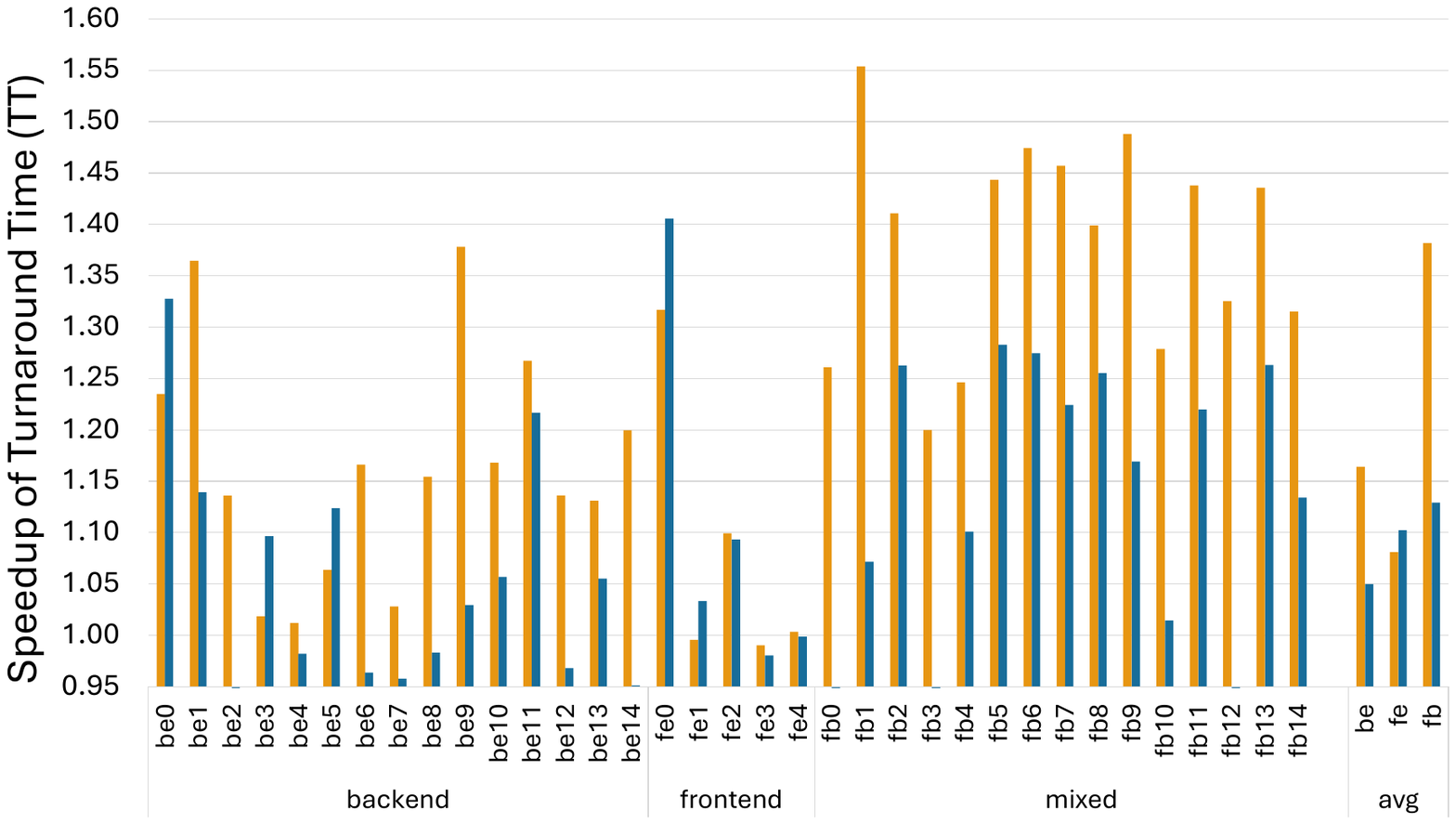}
        \label{fig:TT_synpa_hy}
    }
    \hfill
    \subfloat[Instructions Per Cycle (IPC)]{
        \centering
        \includegraphics[width=\columnwidth, trim = 0 70 0 70, clip]{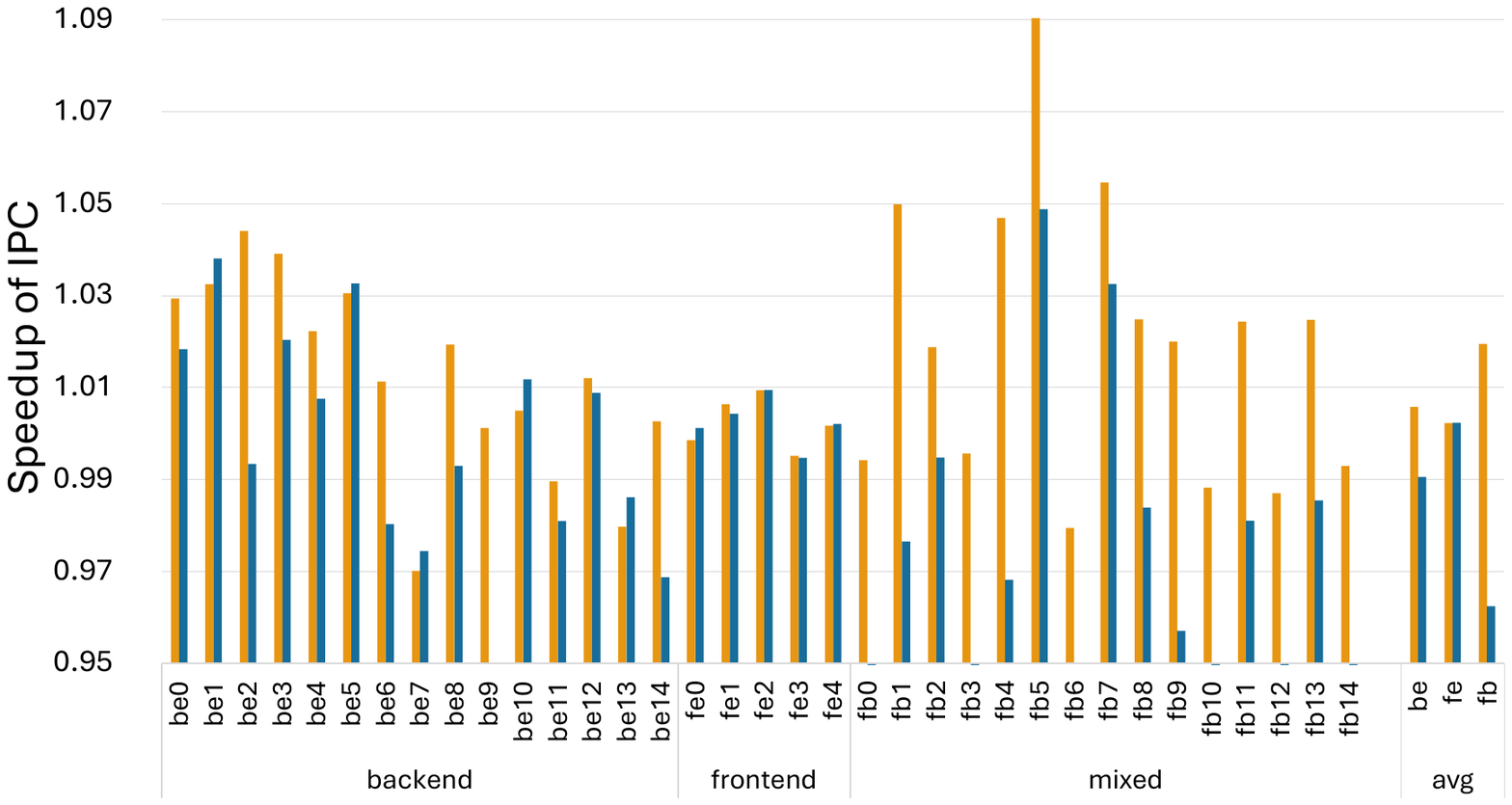}
        \label{fig:IPC_synpa_hy}
    }
    \caption{Speedups of $SYNPA4_{R-FEBE}$ and Hy-Sched.}
    \label{fig:Results_synpa_hy}
\end{figure}

Figure \ref{fig:Results_synpa_hy} compares the performance of $SYNPA4_{R-FEBE}$ and Hy-Sched in terms of TT and IPC. Figure \ref{fig:TT_synpa_hy} shows the speedup of TT achieved by the two policies compared to Linux. The results illustrate that $SYNPA4_{R-FEBE}$ significantly outperforms Hy-Sched in the Mixed workloads. On average, across these workloads, SYNPA outperforms Linux by $38\%$ while Hy-Sched only achieves a speedup of 13\%. This performance difference shows that even though an approach that schedules complementary applications (from the dominant category perspective) in each core significantly outperforms the Linux scheduling policy. A more sophisticated approach, considering more categories, such as $SYNPA4_{R-FEBE}$, which estimates the performance of each possible pair of applications to select the optimal one, has the potential to achieve significantly higher performance benefits. In the Backend-intensive and Frontend-intensive workloads, the performance difference is smaller among both policies than in Mixed workloads. This was expected since the lower the diversity in the behavior of the applications, the less the potential to find schedules that achieve superior performance.

Figure \ref{fig:IPC_synpa_hy} shows that the difference between both policies in terms of IPC is also significant. In most workloads, Hy-Sched performs slightly worse than Linux, especially in Mixed workloads, which is the common case. For instance, in workload \texttt{fb1}, $SYNPA4_{R-FEBE}$ outperforms Linux by 5\% while Hy-Sched performs 3\% worse than Linux. 
In contrast, in a few Frontend-intensive and Backend-intensive workloads, Hy-Sched is able to outperform $SYNPA4_{R-FEBE}$ slightly. The reason is that, as explained above, contrary to the Mixed workload scenario, these types of workloads include mostly applications of the same type. Hence, there are not many possibilities for pairing the applications with an adequate co-runner in the same SMT core. Consequently, both policies struggle to allocate the applications using their corresponding criteria, and both obtain similar results that seem to be as high as the workloads can be improved.  
To sum up, $SYNPA4_{R-FEBE}$ significantly outperforms Linux and Hy-Sched in both TT and IPC, no matter the behavior of the workloads. 

\section{Conclusions}

State-of-the-art T2C allocation policies rely on performance prediction models that are fed by performance stacks gathered at run-time in commercial processors. In this paper, we propose the ISC stack that enhances the accuracy of the prediction model and, consequently, the performance of the T2C allocation policy in the ARM processor.
 
Two main conclusions can be drawn from the experimental results that help maximize the accuracy of the performance prediction model. First, the horizontal waste cycles should be managed as an independent and separate category to build the ISC stack. Second, when the ISC stack height exceeds the execution cycles, the best design choice is to perform a weighted removal of the excess cycles from both the backend and frontend stall categories. 

The discussions carried out throughout this paper are aimed at helping other performance analysts develop their performance stacks in other processors, as the discussed problems can be generalized to any high-performance SMT processor.


\section*{Acknowledgments}
This work has been partially supported by the Spanish Ministerio de Ciencia e Innovación and European ERDF under grants PID2021-123627OB-C51 and TED2021-130233B-C32.
Marta Navarro is supported by Subvenciones para la contrataci\'on de personal investigador predoctoral by CIACIF/2021/413, Josu\'e Feliu  by the RYC2021-030862-I contract, funded by the MCIN/AEI /10.13039/501100011033 and the European Union NextGenerationEU/PRTR.

\bibliographystyle{elsarticle-num} 
\bibliography{biblio.bib}

\end{document}